# DEVELOPMENT OF A NANOSTRUCTUAL MICROWAVE PROBE BASED ON GaAs

*Yang Ju, Tetsuya Kobayashi, Hitoshi Soyama*

*Department of Nanomechanics, Tohoku University*

## ABSTRACT

In order to develop a new structure microwave probe, the fabrication of AFM probe on the GaAs wafer was studied. A waveguide was introduced by evaporating Au film on the top and bottom surfaces of the GaAs AFM probe. A tip having 7 μm high, 2.0 aspect ratio was formed. The dimensions of the cantilever are 250×30×15 μm. The open structure of the waveguide at the tip of the probe was obtained by using FIB fabrication. AFM image and profile analysis for a standard sample obtained by the fabricated GaAs microwave probe and commercial Si AFM probe indicate that the fabricated probe has the similar capability for the measurement of topography of materials.

## 1. INTRODUCTION

With the development of nanotechnology, the measurement of electrical properties in local area of materials and devices has become a great need. Although a lot kinds of scanning probe microscope have been developed for satisfying the requirement of nanotechnology, a microscope technique which can determine electrical properties in local area is still not developed yet. Recently, microwave microscope has been an interest to many researchers [1-4], due to its potential in the evaluation of electrical properties of materials and devices. The advance of microwave is that the response of materials is directly relative to the electromagnetic properties of materials.

In this paper, the development of a nanostructural microwave probe was demonstrated. To restrain the attenuation of microwave in the probe, GaAs was used as the substrate of the probe. The new structural microwave probe is expected to be used for the measurement of the electrical properties as well as the topography of materials and devices.

## 2. PROBE FABRICATION

To obtain the desired structure, wet etching was used to fabricate the probe. Different with the dry etching, a side-etching will occur under the etching mask. Utilizing this property, a cone shaped micro tip can be obtained. Early studies suggested that the square resist pattern having 14 μm sides and one side to be 45° to the <011> direction was found to be the most suitable etching mask for etching the tip of the probe [5]. In the case of single crystalline wafer, such as Si and GaAs, the chemical activities are different for different crystalline planes, thereby, the etch rates are also different. Therefore, the side plane obtained at the side of the mask pattern is the most inactive plane (that is the plane having the most low etching speed) which is parallel to the side of the mask pattern. Consequently, the result of etching is strong affected by the direction of mask pattern [6-8]. By considering the chemical activities at different crystalline planes, the length direction of the etching mask for forming the beam of the cantilever was patterned along the <011> direction. Consequently, the side-etching occurs under the resist mask, and mesa type planes appear at the both sides of the beam (45º inclined plane). On the other hand, inverse-mesa type plane is formed at the end of the beam (60-75º inclined plane). Under the same conditions as the beam fabrication process, holder was formed by back side etching. The etching mask was patterned on the bottom surface, and etching was carried out until the substrate was penetrated. The process to fabricate a GaAs AFM probe was described in details in Ref. [9]. On the other hand, Au film was deposited on top and bottom surfaces of the probe to propagate a microwave signal in the probe. The thickness of the film is 50 nm. Both plane surfaces of the waveguide which were evaporated Au film are connected at the end of the beam. However, there is no Au film on the sides of the beam, since the formed inclined planes at the beam sides are not face to the direction of the evaporation. By using focused ion beam (FIB) fabrication, a slit at the tip of probe was formed to open the connection of the Au film on the two surfaces of the probe.

## 3. PROBE EVALUATION

Fig 1 shows Scanning electron microscopy (SEM) photography of the tip of the fabricated microwave probe based on GaAs. The tip is observed located near the front edge of the cantilever. It is 7 μm high and having an aspect ratio of 2.0. the width of the micro slit is about 200 nm.

In order to confirm the resolution of the fabricated GaAs microwave probes, the AFM topography of two grating samples having 2000 line/mm and 17.9 nm step height were measured by the fabricated GaAs microwave probes, and commercial Si AFM probe, respectively. JSPM-5400 was used for measurement of the sample in non-contact mode. The properties of these AFM probes were given in Table 1. The resonance frequency was swept and Q-value is defined by the following relation, $Q = f_0/(f_+ - f_-)$, where $Q$ is the Q-value of the probe, $f_0$ the scanned peak frequency (resonance frequency), $f_+$ and $f_-$ the shifted frequency from $f_0$ at the 70.7 % of peak intensity. Q- value indicates a resonance sharpness of cantilever, the higher of the Q-value, the better stabilizing of the oscillation. As shown in Table 1, fabricated GaAs microwave probe C and E have higher Q-value than the commercial Si probe.

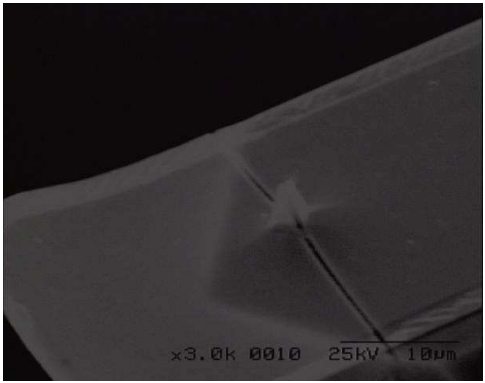

Fig. 1 SEM photograph of the tip and the micro slit of the GaAs microwave probe

### 3.1. Evaluation of topographies

Fig. 2-5 show the non-contact mode AFM topographies of the grating sample having 2000 lines/mm obtained by the fabricated GaAs microwave probe A, C, D, E. The scan area was 3×3 μm. The white spots in these figures are due to micro-dusts on the sample surface. Higher resolution topographies were obtained by probe C and E which have the high Q-value comparing to probe A and D. However, the probe B didn't take good scanning performance because of the lower Q-value.

### 3.2. Comparison with Si AFM probe

Fig. 3 shows the topography of the sample obtained by the fabricated GaAs microwave probe C, which indicates that the grating depth is 20-30 nm. Similar as

Table 1 The properties of AFM probes in the atmosphere

| Probe | The resonance frequency (kHz) | Q-value |
|---|---|---|
| Microwave probe A | 454 | 313 |
| Microwave probe B | 99 | 185 |
| Microwave probe C | 118 | 676 |
| Microwave probe D | 141 | 336 |
| Microwave probe E | 503 | 516 |
| Commercial probe (Si) | 258 | 440 |

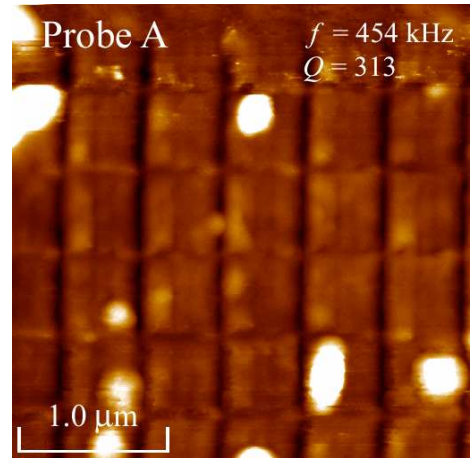

Fig. 2 Surface topography of the grating sample obtained by the fabricated GaAs microwave probe A

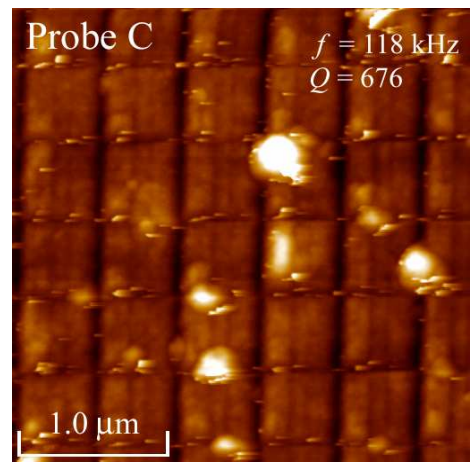

Fig. 3 Surface topography of the grating sample obtained by the fabricated GaAs microwave probe C

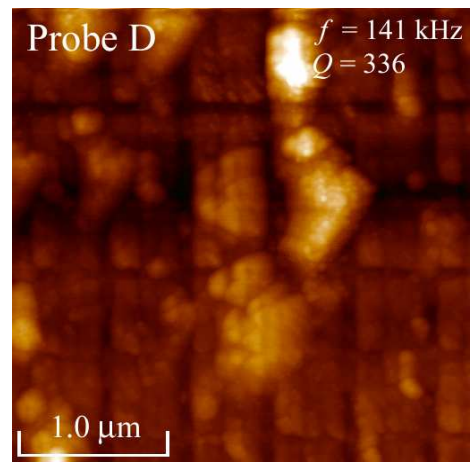

Fig. 4 Surface topography of the grating sample obtained by the fabricated GaAs microwave probe D

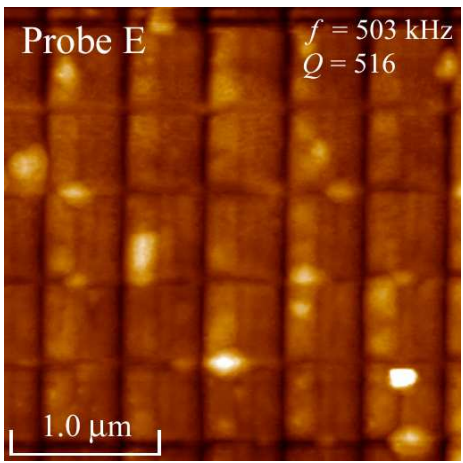

Fig. 5 Surface topography of the grating sample obtained by the fabricated GaAs microwave probe E

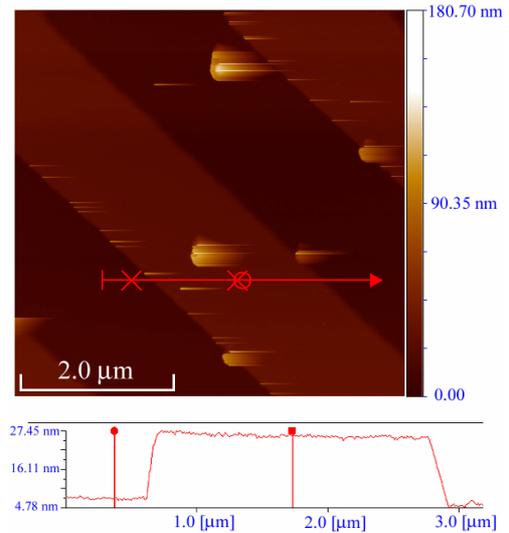

Fig. 7 Profile analysis using by microwave AFM probe C

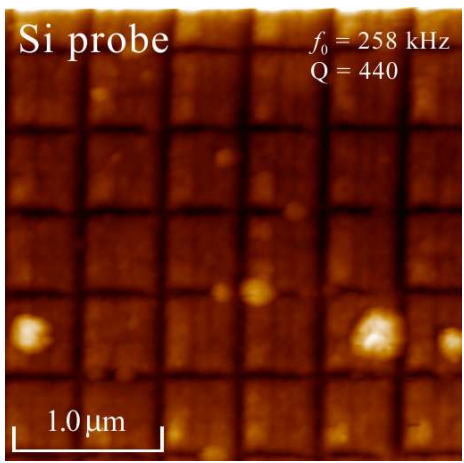

Fig. 6 Surface topography of the grating sample obtained by the commercial Si probe

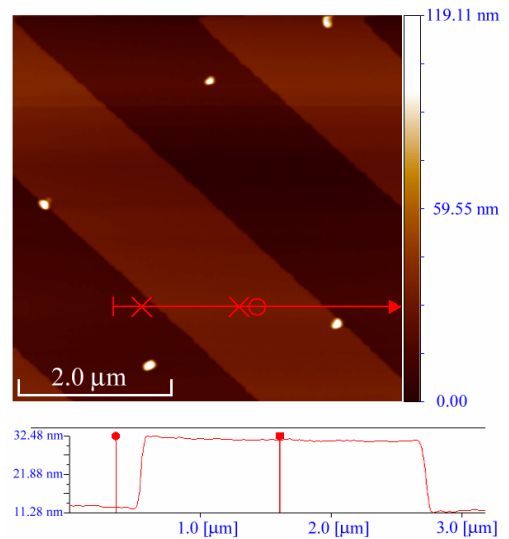

Fig. 8 Profile analysis using by commercial Si AFM probe

probe C, the other GaAs microwave probes also can get AFM topography well. Fig. 6 shows the non-contact mode AFM topography obtained by using commercial Si cantilever, in which the grating depth is shown to be 30-50 nm. Even through the Q-value is lower than that of probe C and E, the commercial Si probe still can obtain a higher resolution topography because of the higher aspect ratio of the tip. These results illustrate that the tip of the GaAs microwave probe having the similar capability to sense the surface topography of materials as that of commercial AFM probe.

### 3.3. Evaluation of height accuracy

In order to evaluate height accuracy, the grating sample having 17.9 nm±1 nm step height was measured by using probe C and Si probe. As shown in Fig. 7, by analyzing profile of the sample, microwave probe C results the step height to be 18.60 nm. In contrast, measurement result by using commercial Si probe indicates the step height to be 18.62 nm, as shown in Fig. 8. These results suggest that the fabricated microwave probe has the similar height evaluation capability as that of commercial AFM probe.

### 4. CONCLUSION

GaAs microwave probes were fabricated on the GaAs wafer by using wet etching process. A waveguide was introducing on the probe by evaporating Au film on the both surfaces of the probe. The open structure of the waveguide at the tip of the probe was obtained by using FIB fabrication. AFM measurements were performed by comparing with the commercial Si AFM probe. It is indicated that GaAs microwave probe has a capability to catch AFM topography of materials and having high accuracy for height evaluation, similar as the commercial AFM probe.

**5. REFERENCES**

[1] Steinhauer, D. E., Vlahacos, C. P. Wellstood, F. C., Anlage, S. M., Canedy, C., Ramesh, R., Stanishevsky, A. and Melngailis, J., "Imaging of microwave permittivity, tenability, and damage recovery in (Ba, Sr) TiO$_3$ thin films," *Applied Physics Letters*, Vol. 75, No. 20, 1999, pp. 3180-3182.

[2] Duewer, F., Gao, C., Takeuchi, I. and Xiang X.-D., "Tip-sample distance feedback control in a scanning evanescent microwave microscope," *Applied Physics Letters*, Vol. 74, No. 18, 1999, pp. 2696-2698.

[3] Tabib-Azar, M., Akinwande, D., "Real-time imaging of semiconductor space-charge regions using high-spatial resolution evanescent microwave microscope," *Review of Scientific Instruments*, Vol. 71, No. 3, 2000, pp. 1460-1465.

[4] Ju, Y., Saka, M. and Abé, H., "NDI of delamination in IC packages using millimeter-waves," *IEEE Transactions on Instrumentation and Measurement*, Vol. 50, No. 4, 2001, pp. 1019-1023.

[5] Ju, Y., Sato, H., and Soyama, H., "Fabrication of the tip of GaAs microwave probe by wet etching", *Proceeding of interPACK2005 (CD-ROM)*, 2005, IPACK2005-73140.

[6] Heisig, S., Danzebrink, H.-U., Leyk, A., Mertin, W., Münster, S. and Oesterschulze, E., "Monolithic gallium arsenide cantilever for scanning near-field microscopy," *Ultramicroscopy*, Vol. 71, No. 1-4, 1998, pp. 99-105.

[7] Iwata, N. Wakayama, T. and Yamada, S., "Establishment of basic process to fabricate full GaAs cantilever for scanning probe microscope applications," *Sensors and Actuators A*, Vol. 111, No. 1, 2004, pp. 26-31.

[8] MacFadyen, D. N., "On the preferential etching of GaAs by $H_2SO_4$-$H_2O_2$-$H_2O$," *Journal of Electrochemical Society*, Vol. 130, No. 9, 1983, pp. 1934-1941.

[9] Ju, Y., Kobayashi, T., and Soyama, H., "Fabrication of a GaAs microwave probe used for atomic force microscope", *Proceeding of interPACK2007 (CD-ROM)*, 2007, IPACK2007-33613, in press.